\documentstyle[preprint,tighten,aps,epsfig]{revtex}

\def\lapprox{\mathrel{\mathop  {\hbox{\lower0.5ex\hbox{$\sim$}
\kern-0.8em\lower-0.7ex\hbox{$<$}}}}}  

\def\gapprox{\mathrel{\mathop  {\hbox{\lower0.5ex\hbox{$\sim$}
\kern-0.8em\lower-0.7ex\hbox{$>$}}}}}

\begin{document}

\draft

\preprint{\vbox{\noindent{}\hfill INFNFE-12-02}}

\title{Neutrinos and Energetics of the Earth}

\author{G.~Fiorentini$^{(1,2)}$, F. Mantovani$^{(3)}$
and B. Ricci $^{(1,2)}$}

\address{
$^{(1)}$Dipartimento di Fisica dell'Universit\`a di Ferrara, I-44100
Ferrara, Italy\\
$^{(2)}$Istituto Nazionale di Fisica Nucleare, Sezione di Ferrara, 
I-44100 Ferrara, Italy \\
$^{(3)}$Scuola di Dottorato, Dipartimento di Scienze della Terra, Universit\'{a} di Siena, 
53100 Siena, Italy
}

\date{\today}

\maketitle

\begin{abstract}
We estimate terrestrial antineutrino and neutrino fluxes according to           
different models of Earth composition.                                          
We find large variations, corresponding to uncertainties on the                 
estimated $U$, $Th$ and $K$ abundances in the mantle.                                 
Information on the mantle composition can be derived from antineutrino          
flux measurements after subtracting the crust contribution. This                
requires a good description of the crust composition in the region of           
the detector site.   Measurements of terrestrial antineutrinos will             
provide a direct insight   on the  main sources of Earth's heat flow.           
                                               
\end{abstract}

\section{Introduction}
Earth emits a tiny heat flux with an average value $\Phi_H= 80 \,mW/m^2$, 
definitely smaller than the radiation coming from the Sun, $ K_{\odot} =1.4\, 
kW/m^2$, larger however than the energy deposited by cosmic rays, $\Phi_c \simeq 
10^{-8}W/m^2$. When integrated over the Earth surface, the tiny flux translates 
into a huge heat flow, $H_{\oplus} \simeq 40 \, TW$, the equivalent of ten thousand 
nuclear power plants \cite{Bro}.

We would like to recall to  the particle physics community that 
the sources of Earth energy flow are not understood quantitatively and that 
measurements of (anti)neutrinos from the Earth in the next few years should be 
capable of determining the radiogenic contribution.  

A comparison between the Sun and Earth energy inventories may be useful for 
illustrating the differences in the two cases. Clearly, a heat flow $H$ can be 
sustained for a time $t$ provided that an energy source of at least $U=H\,t$ is 
available.

For the Sun, $U=H_\odot \,t_\odot\simeq 5 \cdot 10^{43}J$ 
and clearly  neither gravitation ($U_G \simeq 
GM^2_\odot/R_{\odot}=4\cdot 10^{41} \,J$  ) nor chemical reactions ($U_{ch} 
\simeq 0.1\, eV \cdot N_\odot= 2 \cdot 10^{37}\, J$, where $N_{\odot}$ is the number of 
nucleons) are  enough, and only nuclear energy ($U_{nuc} \simeq 1\,MeV \,N_\odot= 2 
\cdot 10^{44} \, J$) can sustain the solar luminosity over the solar age, as 
beautifully demonstrated  Gallium experiments in the last decade \cite{Gal}. On the other 
hand for the Earth one has $U_G \simeq 4 \cdot 10^{32} \,J $, $U_{ch} \simeq 6 \cdot 
10^{31} \, J$ and $U_{nuc} \simeq  6 \cdot 10^{30} \, J$ (assuming some that 
some $10^{-8}$ of Earth mass  consists of radioactive nuclei), so that  each of 
the previous  mechanisms   in principle can  account for $U_\oplus= 5 \cdot 10^{30}\, 
J$. In order to understand the energetics of the Earth one has to clarify the 
roles of the different energy sources, their locations and when they have been 
at work. At the end of a review on the Earth energy sources 
J. Verhoogen \cite{Ver} summarized the situation with the following words:
{\em ``What emerges from this morass of fragmentary and uncertain data is that 
radioactivity by itself could plausibly account for at least 60 percent, if not 
100 percent, of the earth's heat output. If one adds the greater rate of 
radiogenic heat production in the past, ... possible release of gravitational 
energy (original heat, separation of core, separation of inner core, tidal 
friction,... meteoritic impact ...), the total supply of energy may seem 
embarrassingly large. ... Most, if not all of the figures mentioned above  are 
uncertain by a factor of at least 2, so that disentangling contributions from 
the several sources is not an easy problem.''}. 

In this respect a determination 
of the radiogenic contribution is most important.
Radiogenic heat arises  mainly\footnote{For simplicity 
we neglect $^{235}U$ and $^{87}Rb$ which provide 
smaller contributions.}
from the decay (chains) of $^{238}U$, $^{232}Th$ and $^{40}K$. 
All these elements produce heat together with antineutrinos, with well fixed 
ratios heat/neutrinos. A measurement of the antineutrino flux, and possibly of 
the spectrum, would provide a direct information on the amount and composition 
of radioactive material inside Earth and thus would determine the radiogenic 
contribution to the heat flow. 

On the other hand, until recently the neutrino 
fate could not be predicted reliably, as testified by the thirty years old solar 
neutrino puzzle \cite{Bah}. The disagreement between theory and observation by factors of 
order two suggested that (anti)neutrino survival probabilities were essentially known 
within factors of two. Thus observation of terrestrial (anti)neutrinos could not be 
useful for improving our knowledge of Earth radioactivity. The situation has 
dramatically changed since the SNO results \cite{Sno}, which clearly prove that a fraction  
of electron neutrinos  change their flavor during the trip form Sun to Earth. 
When combined with the results of other solar and terrestrial neutrino 
experiments, the picture is converging towards the so called large mixing angle 
(LMA) oscillation solution. In other words, now we can predict reliably the fate of 
terrestrial neutrinos and antineutrinos, in their trip from production site to 
detectors.

Last but not least, the experimental techniques for detection of MeV 
antineutrinos have enormously improved in the last few years. As testified by the 
development of Kamland \cite{Kam} and Borexino \cite{Bor}, it is now possible to build kiloton size 
detectors, with extremely low background.

The argument of  geo-neutrinos was introduced by Eder \cite{Eder} in the sixties, it
was reviewed extensively by Krauss, Glashow and Schramm \cite{Krauss} in the eighties
and it has been considered more recently in \cite{Rag,Rot}. Now it is the right time
for neutrino physics to contribute in reconstructing the thermal history
of the Earth.

\section{Energy sources and neutrino luminosities}
\label{energy}

The heat production rates  per unit mass of \underline{natural} U, Th and K are given 
by\footnote{The marked difference of $\epsilon (K)$ corresponds to the fact that 
the natural abundance of $^{40}K$ is $1.2 \cdot 10^{-4}$, i.e. $\epsilon (^{40}K) = 
0.3\cdot 10^{4} \,W/kg$.}:

\begin{eqnarray}
\label{epsi1}
\epsilon(U)&=& 0.95 \cdot 10^{-4}\quad {\em W/kg} \\ \nonumber
\epsilon(Th)&=& 0.27 \cdot 10^{-4} \quad {\em W/kg} \\ \nonumber
\epsilon(K)&=& 0.36 \cdot 10^{-8} \quad {\em W/kg}\\ \nonumber
\end{eqnarray}

This is sufficient to determine the Earth radiogenic heat production rate $H$ 
in terms of the mass of each element. When heat production is expressed in TW 
and masses in units of $10^{17}\, kg$ one has:

\begin{equation}
\label{lum1}
H=9.5\,M(U)+2.7\,M(Th)+3.6\cdot 10^{-4}\,M(K)
\end{equation}

It is convenient to write this equation in terms of the uranium mass $M(U)$ and 
of the mass ratios of the other elements to $U$, as these latter quantities are  
more regularly distributed  in terrestrial and meteoritic samples:

\begin{equation}
\label{lum2}
H= 9.5 \,M(U)\,[ 1+ 0.28 \, Th/U + 3.8 \cdot 10^{-5} \, K/U ]
\end{equation}

The specific neutrino production rate (neutrinos per unit mass and time) of each 
element  $\epsilon_\nu $, is immediately derived from the isotopic abundance, 
decay time and the number of neutrinos emitted in each decay, see Table \ref{Tabdecay}.
(Anti)Neutrinos luminosities are immediately derived in terms of the mass of each element
and the appropriate $\epsilon_{\bar{\nu} / \nu}$.
Measuring $L_{\bar{\nu}/ \nu }$ in units of $10^{24}$ particles per second and 
masses in units of $10^{17}\,kg$ one has:
\begin{eqnarray}
\label{lnu1}
L_{\bar{\nu}}&=& 7.4 \, M(U)+1.6 \, M(Th)+2.7\cdot 10^{-3} \, M(K) \\ \nonumber
L_\nu&=&3.3 \cdot 10^{-4} \,M(K)\\
\end{eqnarray}

We have thus the basic equations for determining radiogenic heat production and 
neutrino flows from models of the Earth composition.

\subsection{A naive chondritic earth}
\label{seccho}

The simplest model assumes that the global composition of the Earth is similar 
to that of the oldest meteorites, the carbonaceous chondrites (CI).

The  typical values  of CI  \cite{And} are $Th/U=3.8$ ,   $K/U =7\cdot 10^4$  and  
$U/Si=7.3\cdot 10^{-8}$ \cite{Bro}. Silicon represents about 15\% of the Earth mass, 
$M_\oplus=5.97\cdot 10^{24}\,kg$. If these elements have not been lost in the 
Earth formation process,  one obtains 
$M(U)=0.653\cdot 10^{17}\, kg$, 
$M(Th)=2.48\cdot 10^{17}\, kg$ and 
$M(K)=4.57\cdot 10^{21} \,kg$.

The contribution to heat flows and neutrino luminosities are reported in the 
first column of  Table \ref{Tabenergy}. Radiogenic production in the chondritic 
model  easily accounts for 75\% of the observed heat flow, and it could easily 
saturate it when uncertainties are included. Uranium and thorium provide 
comparable contributions, each a factor of two below that of potassium. 
Concerning antineutrinos, potassium dominates by an order of magnitude at least, 
as a consequence of the more favourable neutrino/energy ratio.

\subsection{ The Bulk Silicate Earth  model}
\label{secbse}

Uranium, thorium and potassium are lithophile elements, so  they should
accumulate in the Earth crust. Actually, upon averaging data over the huge
differences between continental and oceanic components, it is found that
Earth crust contains some $3/4$ of the  uranium predicted for the whole Earth
by the chondritic model \cite{Bro}. Within large variations, $Th/U$ is consistent with
the chondritic prediction. On the other hand, the crust looks depleted in
potassium, the typical ratio being $K/U=10,000$,  a factor 7 below that of
CI.

Observational data on the mantle, which are anyhow limited to the upper
part,
suggest that uranium and potassium are globally more abundant than the CI
prediction, $Th/U$ is consistent with the chondritic value and 
the potassium depletion is confirmed. No observational data are available on the core,
which should consist of siderophile elements without significant amount of 
$U, \, Th$ or $K$.

Actually, when deriving  Earth composition from  meteoritic data, one has
to take into account the volatilization   of  a significant fraction (some
17\%) \cite{Rin} of the total $SiO_2$, so that a larger amount of metoritic material
is needed for Earth formation. 

The origin of potassium depletion, also observed in the Moon, Venus and
Martian meteorites, is somehow uncertain. Elements of the atomic weight of
potassium cannot be lost from the terrestrial planets, even at elevated
temperatures, once these bodies have reached their present size \cite{Und}. The
most reasonable  explanations seems that this element was depleted in the
precursor planetesimals from which the inner planets 
accumulated\footnote{ It has been suggested that potassium behaves as a metal at high
pressure, and thus it can be buried in the planetary cores. This
hypothesis could work for Earth, and it provides a suitably placed energy
source for sustaining the terrestrial  magnetic field, see \cite{Bro}. However it does not
explain potassium depletion in Mars, where the central pressure, only 400
kbars,  is insufficient for  potassium to enter a Martian core.}.

All this brings us to the Bulk Silicate Earth (BSE) model, which  provides
a description of  geological evidence  coherent with  geochemical
information.
It describes the primordial mantle, prior to crust separation. The estimated 
uranium mass is:
\begin{equation}
\label{mutot}
M(U)=0.8\cdot 10^{17}\, kg \quad,
\end{equation} 
within some 20\% \cite{Mc},  the ratio $Th/U$ is
close to the chondritic value and $K/U=10,000$. The present crust and mantle
should contain respectively about one half of each element.

In this BSE model   the (present) radiogenic production, mainly from uranium and thorium,
accounts for about one half  of the total heat flow.
The antineutrino luminosities  from uranium and thorium are rescaled by a factor
1.3  whereas potassium, although reduced by a factor of 5, is still the
principal antineutrino source.

\subsection{ A fully radiogenic model}
\label{secfully}

At the other extreme, one can conceive a model where heat production is
fully radiogenic, with $K/U$ fixed at the terrestrial value and  $Th/U$ at the
chondritic value, which seems consistent with terrestrial observations.
All the abundances are rescaled so as to provide the full $40 \,TW$ heat
flow (last column of table \ref{Tabenergy}). 
All particle production rates  are correspondingly
re-scaled by a factor of two with respect to the predictions of the BSE
model.

\vspace{1cm}

In summary, the discussion of these somehow extreme models shows that particle 
luminosities are uncertain  by a factor of order two, the relative  contributions 
to heat production are  strongly model dependent whereas potassium is anyhow the 
principal neutrino and anti-neutrino source.

\section{From luminosity to flux and signal}

An (anti)neutrino detector near the Earth surface ($R=R_\oplus$) is sensitive to the  flux 
impinging onto it from any direction: 
\begin{equation}
\label{phi}
\Phi = \frac{1}{4 \pi} \int d^3r \frac{A(\vec{r})} {|\vec{R}-\vec{r}|^2} \quad , 
\end{equation}
where the integral is taken  over the Earth volume and  $A $ is the number of 
particles produced per unit volume and 
time\footnote{
This is different from the flux normal to the Earth surface, which in 
the case of spherical symmetry is given by $\Phi_\perp = \frac {1}{4 \pi R_\oplus^2} \int d^3r A(r)$.}.
The flux  depends on the geometrical distribution of the sources, and we  we can 
write: 
\begin{equation}
\label{phi2}
\Phi_{\bar{\nu}/\nu}= \frac{G L_{\bar{\nu}/\nu}} { 4\pi R_\oplus^2}
\end{equation}
where $ G $ is a geometrical factor of order unity. One has:
$ \Phi/(10^6\,cm^{-1}\,s^{-1})=0.2 \, G \, L/ ( 10^{24}\,s^{-1} )$.

In order to estimate fluxes, one  needs to know the distribution of  radioactive elements in the Earth interior, 
and not  only their total abundances.
Concerning the crust, these elements are mainly concentrated in the continental part. Recent estimates of the uranium average mass
abundance in the continental crust $(CC)$ are near $a_{CC}(U) =1.7 \,ppm $ \cite{Wedepohl}. 
The abundance in the oceanic crust  $(OC)$ is an order of magnitude smaller, $a_{OC}\simeq 0.1\, ppm$. 
If one also considers that  $OC$ is much thinner than $CC$  ($ M_{CC} =2.3 \cdot 10^{22}\,  kg$  and  
$M_{OC}= 0.6\cdot 10^{22}\, kg $)
one  concludes that the contribution of the oceanic crust  to neutrino production is definitely smaller. 
In this way we estimate the total uranium mass in the crust:
\begin{equation}
\label{mucrust} 
		M_{c}(U)= 0.4 \cdot 10^{17} \, kg \quad .
\end{equation}

For the sake of a first estimate, we shall assume that  $M_{c}(U)$ is uniformly distributed over the 
Earth surface  within a layer of thickness $\bar{h}=30 \,km$.
We associate to the mantle  an uranium mass $M_m (U) = M(U)- M_{c}(U)$, where $M(U)$
 was  estimated in the
 previous section for each model and $ M_{c}(U)$ is given from eq.(\ref{mucrust}). 
Furthermore, we shall assume uniform distribution within the mantle. We also assume a 
uniform distribution of $Th/U$ and $K/U$. 
Within these approximations the geometrical factors $G$ are easily calculated.

For a spherical shell with radii $r_1=x_1 R_\oplus$ and $r_2=x_2 R_\oplus$ and uniform 
distribution one has:

\begin{equation}
\label{gfactor}
G=\frac{3}{2(x_2^3 -x_1^3)}  \left \{ \frac{1}{2} (1-x_1^2) \ln \frac{1+x_1}{1-x_1} -x_1
                                   -\frac{1}{2} (1-x_2^2) \ln \frac{1+x_2}{1-x_2} +x_2  \right \}
\end{equation}

For the crust ($r_2=R_\oplus$ and $r_2-r_1=\bar{h} \simeq 30\, km$) and for the mantle 
($r_2\simeq R_\oplus$ and $r_1\simeq R_\oplus/2$) one has:
\begin{equation}
\label{gfactor2}
G_c \simeq \frac{1}{2} [1+ln(2R_\oplus/\bar{h})] = 3.5  \quad ; \quad  G_m=1.6 
\end{equation}

This allows the calculation of the fluxes for each Earth model reported in Table 
\ref{Tabflussi}.

One remarks that fluxes  are of the order of magnitude of the solar Boron neutrino
flux. potassium (anti)neutrinos are the dominant component in any model.
The various models yield significantly different predictions. Contributions
from crust and mantle look comparable.

Indeed in order to reach these results we used several approximations, which is worth discussing, also
in view of obtaining more precise estimates.

i){\em Global effects}.\\
There are significant uncertainties on the value of $a_{cc}(U)$. 
For the lowest value  found in the literature, $a_{CC}(U) =0.91 \, ppm $ \cite{MacLennan85}, 
the  uranium mass in the crust is halved with respect to the reference value  in eq.(\ref{mucrust}). 
In addition, there  are  indications that the upper part of the mantle
 is impoverished in the content of radioactive elements.
 The effect of these uncertainties are shown in Table \ref{Tabhigh}, where $\Phi_{\bar{\nu}}(U)$
 is calculated  for a fixed (BSE) value of the crust+mantle uranium mass.  
A change of the  $U$-abundance in the crust by a factor 
of two corresponds  to  a 15\%  change in the flux. If  uranium is completely removed from the upper mantle  
the flux is reduced by 3\%. All these effects are   thus smaller with respect to the uncertainty on the total 
 uranium mass, which  is reflected in changes of fluxes by factor two when going through the extreme 
chondritic and fully radiogenic models.
Similar considerations hold for thorium and potassium

ii){\em Regional effects}.\\
The fluxes reported in Table \ref{Tabflussi} are averaged values and 
one has to remind that Earth crust is significantly variable 
in thickness and composition.
Thus one has to expect  significant 
variations of the actual fluxes, depending on the detector location.

As a few significant examples, the Gran Sasso laboratory sits on a thick continental crust, 
whereas Kamioka is on an island arc in between the Eurasian plate and the Pacific plate. 
Following the approach of ref. \cite{Rot} we have estimated $\Phi_{\bar{\nu}/\nu}$
at the two sites,  in a model which   
distinguishes between $CC$ and  $OC$ and includes  a variable crust thickness. The model is based over 
a global crustal map at $5^o \times 5^o$, and assumes uniform distribution within the mantle. Uranium mass  
in the crust and in  total Earth  are kept fixed at the values of eqs. (\ref{mutot}) and (\ref{mucrust}).
Table \ref{Tabsites} shows that fluxes at 
specific sites can  differ significantly  from the average value and thus a crust map is really 
needed for a precise flux estimate at the detector size. On the other hand,  the difference between 
Gran Sasso and Kamioka is at the level of 15\%.

iii){\em Local effects.}\\
The flux  from the crust within a distance $d$ from the detector is easily 
estimated using planar geometry $(d<<R_\oplus)$:
\begin{equation}
\label{phid}
\Phi(<d) =  \frac{A}{2} [d \, \arctan (h/d) + (h/2)ln(1+d^2/h^2)]
\end{equation}
where $h$ is the local crust thickness and $A$ is the local activity.
This has to be compared with the total flux 
from the crust:
\begin{equation}
\label{phic}
\Phi_c = G_c \bar{A} \bar{h}.
\end{equation}
where $\bar{A}$ and $ \bar{h}$ are the mean crustal activity and thickness.

A significant quantity is the  relative contribution $R= \Phi(<d)/\Phi_c$ . By 
suitable expansions  of eq. (\ref{phid}) one  immediately derives the 
contribution of the nearby rocks (say $d=h/3$) and of the local area, i.e. up 
to a distance $d<<R_\oplus$:
\begin{eqnarray}
\label{rnea}
R_{nea} &=&  0.07 Ah/(\bar{A}\bar{h}) \\ 
R_{loc} &=&  (1/7)[1+\ln (d/h)] Ah/(\bar{A}\bar{h})  \quad .\\
\end{eqnarray}

If $A=\bar{A}$ and $h=\bar{h}=30$ km one finds that  the  rocks within 10 $km$ 
contribute 7\% with respect to the total from the crust. Within 100 $km$ 
one has about  30\%  of the crust contribution and about 15\% of the total flow,
from crust and below. 
A more detailed  geochemical information of this area is thus needed if one aims at a few percent accuracy

In summary, detection of terrestrial (anti)neutrinos is particularly important for 
determining the amount of radioactive material inside the mantle, where 
information on the chemical composition is most uncertain, the lower part being 
completely unaccessible to observation.  A suitable approach would thus consist 
of i) measuring the (anti)neutrino flux; ii) subtracting  the component originated 
in the crust, which has been mapped with geological methods, so as to  determine  
the corresponding abundances in the mantle.
The previous calculations show that crust and mantle  provide comparable fluxes, 
so that the subtraction procedure is possible.
The crust 
description however  has to be more detailed in the proximity of the detector.

A detailed discussion of the (anti)neutrino signal is beyond the aim of this 
letter and we would like to remark just a few relevant points.

i)Due to the different antineutrino energy end-points \cite{Eder} ($E_{max}$= 3.26, 2.25 and 
1.31 MeV for $U$, $Th$ and $^{40}K$ respectively) it is possible at least in 
principle to separate the contributions to Earth radioactivity.

ii)Antineutrinos from $U$ and $Th$ can be detected and separated by means of
\begin{equation}
\label{eqnu}
\bar{\nu}+ p \rightarrow e^+ + n - 1.804 \,MeV \quad ,
\end{equation}
whereas different detection schemes are necessary for $K$ antineutrinos, which are below 
the energy threshold for (\ref{eqnu}).  The monochromatic ($E=1.513 \,MeV$  ) neutrinos from $^{40}K$ are 
essentially obscured by  the dominant solar flux ($\Phi(pep)= 1.4 \cdot 10^8 \, cm^{-2}\, s^{-1}$ 
at $E =1.44\, MeV$ 
and $\Phi({\small CNO}) \sim 10^9 \,cm^{-2}\, s^{-1}\, MeV^{-1}) $ .

iii)So far we did not consider the effect of neutrino oscillations. If LMA is 
the correct solution, with $\Delta m^2\simeq 5.5 \cdot 10^{-5}\, eV^2$  
and $\sin ^2 2\theta \simeq 0.83$ \cite{Fogli},
the oscillation length $L=4 \pi E/\Delta m^2$
of 1 MeV antineutrino is around 45 km and  the 
(distance averaged) survival probability of electron antineutrinos  $P_{ee} = 1-
1/2 \sin ^2 2\theta\simeq 0.58$ 
The present uncertainty on $\sin ^2 2 \theta$ (about 20\%) translates into a 15\% uncertainty on the 
fluxes.

iv) Events from $U$ and $Th$ antineutrinos in a organic scintillator detector
have been estimated in the range (20-100)/kton-year \cite{Rag,Rot},
 so that a flux measurement  with a 10\% accuracy should be feasible in a few 
years.
The main background source \cite{Lag} is antineutrinos from nuclear power plants (see last 
row of Table \ref{Tabflussi}), which depends on the detector location.

\section{Concluding remarks}

We have estimated terrestrial antineutrino and neutrino fluxes according        
to different models of Earth composition. We find large variations,             
corresponding to uncertainties on the estimated $U$, $Th$ and $K$ abundances          
in the mantle.                                                                  
Information on the mantle composition can thus be derived from                  
(anti)neutrino flux measurements after subtracting the crust                    
contribution. This requires a good description of the crust composition         
in the region of  the detector site and in return  it will provide              
direct insight  on the main sources of Earth's heat flow.           

Just a few years after the celebrated slow neutron studies of the Rome group, 
Bruno Pontecorvo developed the neutron well log \cite{Ponte}, an instrument  which is still 
used in geology for the search and analysis of hydrogen containing substances 
(water and hydrocarbons). Possibly it is now the time for applying to different 
disciplines what we have learnt so far on neutrinos. In fact, there are several 
attempts in this direction, see e.g.  \cite{Smirnov} and references therein.
The determination of the radiogenic 
component of the terrestrial heat is an important and so far unanswered 
question. It  looks to us  as  the first fruit which we can get from neutrinos,
and Kamland will catch the firstlings very soon.

\vspace{2cm}

{\bf Aknowledgment}
It is a pleasure to acknowledge several interesting discussions with 
with  L. Beccaluva, G. Bellini, E. Bellotti, C. Bonadiman, C. Broggini,  L.
Carmignani, M. Lissia, G. Ottonello, S. Schoenert, C. Vaccaro and F.L. Villante.

\vspace{5cm}
{\bf Addendum}\\
After this paper was submitted, the first results of KamLAND have become available \cite{kamland}.
From an exposure of $1.39 \cdot 10^{31}$ protons$ \cdot$year,  9
geo-neutrino events are reported.  For the best fit  survival  probability
$P(\bar{\nu}_e \rightarrow \bar{\nu}_e)=0.55$ \cite{kamland},  from the average fluxes
calculated in Table \ref{Tabflussi} one predicts 2.6, 3.1 and 5.1 events
respectively for the
chondritic, BSE and fully-radiogenic models. Predictions from fluxes
calculated specifically for the  Kamioka site are 3.5, 4.0 and 6.0 events
respectively.   They are all consistent with the experimental value,
within its statistical fluctuation of  $\pm 5.7$ events \cite{noi2}.


\begin{table}
\caption[aaa]{{\bf Main radiogenic sources}.
 We report the Q-values, the half lives ($\tau_{1/2}$), the 
maximal energies $(E_{max})$  and  (anti)neutrino production rates $(\epsilon_{\bar{\nu}/ \nu})$ 
per unit mass for \underline{natural} isotopic abundances.
Neutrinos from electron capture are monochromatic. }
\begin{tabular}{lllll}
decay             &     $Q$    &  $\tau_{1/2}$  &    $E_{max}$   &  $\epsilon_{\bar{\nu}/ \nu }  $ \\
                  &    $[MeV]$ &  $[10^9 \,y]$   &    $[MeV]$     &  $[kg^{-1} \, s^{-1}$ \\ 
\hline    
$^{238}U \rightarrow ^{206}Pb +8 ^4He + 6 e +6 \bar{\nu}$ &
 			51.7   &   4.47   &     3.26      &    $7.41\cdot 10^7$  \\
$^{232}Th \rightarrow ^{208}Pb +6 ^4He + 4 e + 4 \bar{\nu}$&
                        42.8   &   14.0   &     2.25      &    $1.63\cdot 10^7$  \\
$^{40}K \rightarrow ^{40}Ca +e +\bar{\nu}$ &
                       1.321   &   1.28   &     1.31      &     $2.69 \cdot 10^4 $\\
$^{40}K + e \rightarrow ^{40}Ar +\nu $ &
                      1.513   &          &      1.51            &     $3.33 \cdot 10^3 $\\
\end{tabular}
\label{Tabdecay}
\end{table}

\begin{table}
\caption[a]{{\bf Masses, heat and neutrino production rates}.
$M, \, H$ and $L$ are in units of
$10^{17} \, kg$, $10^{12} \, W$ and $10^{24} s^{-1}$ respectively. $H_{NR}$ is defined 
as the difference between total heat flow $H_\oplus$ and the radiogenic production.}   
\begin{tabular}{llll}
Model & Chondritic & BSE & Fully Radiogenic\\
\hline
$M(U) $ & 0.65 & 0.84  &1.7\\
$M(Th)$ & 2.5  & 3.4  &6.5\\
$M(K) $ & $4.6 \cdot 10^4$ & $0.84 \cdot 10^4$ & $1.7 \cdot 10^4$ \\
\hline
$H(U) $ & 6.2 & 7.9 & 16.3\\
$H(Th)$ & 6.7 & 8.6 & 17.6\\
$H(K) $ &16.4 & 3.0 &  6.1\\
$H_{NR}$ &10.7 &20.5 & 0\\
\hline
$L_{\bar{\nu}} (U)$ &  4.8  & 6.2 & 12.7 \\
$L_{\bar{\nu}} (Th)$&  4.0  & 5.2 & 10.6\\
$L_{\bar{\nu}} (K)$ &  123   & 22.5 & 46.0 \\
$L_{\nu} (K)$       &  15.2   & 2.8  & 5.7 \\
\end{tabular}
\label{Tabenergy}
\end{table}

\begin{table}
\caption[aa]{{\bf Mass distribution and (anti)neutrino fluxes}.
 $M$ and $\Phi$ are in in units of $10^{17} \, kg$, and $10^6\,cm^{-2} \,s^{-1}$ respectively.
Uranium mass in the crust is fixed at the  value estimated from ref. \cite{Wedepohl}. 
$\Phi_{\bar{\nu}} $(reactor) corresponds to the
flux from a nuclear reactor with $ P_{th}=2.8 \, GW $ at $ 100 \, km$.}
\begin{tabular}{llll}
& \multicolumn{2}{c}{{\bf Uranium}}& \\
\hline
Model & Chondritic & BSE & Fully Radiogenic \\
\hline
$M$(crust)  & 0.42 & 0.42 & 0.42 \\
$M$(mantle) & 0.23 & 0.42 & 1.29  \\
\hline
$\Phi_{\bar{\nu}}$ (crust)  & 2.1  & 2.1  & 2.1 \\
$\Phi_{\bar{\nu}}$ (mantle) & 0.5 &  1.0  & 3.0 \\ 
${\mathbf \Phi_{\bar{\nu}}}$ {\bf (tot)}    & {\bf 2.6}  & {\bf 3.1}  & {\bf 5.1} \\
\hline
\hline
$\Phi_{\bar{\nu}}$(reactor) ($E\leq3.26$ MeV) &  &  0.4 &  \\
\end{tabular}

\vspace{1cm}

\begin{tabular}{llll}
& \multicolumn{2}{c}{{\bf Thorium}}& \\
\hline
Model & Chondritic & BSE & Fully Radiogenic \\
\hline
$M$(crust)  & 1.6  & 1.6  & 1.6  \\
$M$(mantle) & 0.89 & 1.6  & 4.9  \\
\hline
$\Phi_{\bar{\nu}}$ (crust)  & 1.8  & 1.8   & 1.8 \\ 
$\Phi_{\bar{\nu}}$ (mantle) & 0.4  & 0.8   & 2.5 \\
$ {\mathbf \Phi_{\bar{\nu}}}$ {\bf (tot)}    & {\bf 2.2}  & {\bf 2.6}   & {\bf 4.3}  \\
\hline
\hline
 $\Phi_{\bar{\nu}}$(reactor) ($E\leq2.25$ MeV) &  & 0.3 &  \\
\end{tabular}

\vspace{1cm}

\begin{tabular}{llll}
& \multicolumn{2}{c}{{\bf Potassium}}& \\
\hline
Model & Chondritic & BSE & Fully Radiogenic \\
\hline
$M$(crust)$/10^4$  & 0.42  &  0.42  & 0.42 \\
$M$(mantle)$/10^4$ & 4.15   & 0.42   & 1.29 \\
\hline
$\Phi_{\bar{\nu}}$ (crust)  &  7.7  &  7.7   &   7.7  \\
$\Phi_{\bar{\nu}}$ (mantle) & 34.9  &  3.5    & 10.9 \\
${\mathbf \Phi_{\bar{\nu}}}$ {\bf (tot)}    & {\bf 42.6}  &  {\bf 11.2}   & {\bf 18.6}  \\ 
\hline
$\Phi_\nu$ (crust)  & 0.95  &  0.95  & 0.95  \\
$\Phi_\nu$ (mantle) & 4.33  &  0.44  & 1.04  \\
${\mathbf \Phi_\nu }$ {\bf (tot)}    & {\bf 5.28}  &  {\bf 1.39}  & {\bf 1.99}  \\
\hline
\hline
$\Phi_{\bar{\nu}}$(reactor) ($E\leq 1.31$ MeV) &  & 0.2 &  \\
\end{tabular}
\label{Tabflussi}
\end{table}

\begin{table}
\caption[xxx]{{\bf Flux dependence  on uranium abundance in the crust and distribution 
in the mantle.} Fluxes are calculated for $M_{crust+mantle}(U)=0.84\cdot 10^{17} \, kg$.
The total flux $\Phi$(tot) is the sum of the contribution from the crust, upper mantle and
transition zone+lower mantle. Same units as in Table \ref{Tabflussi}.}
\begin{tabular}{llll}
\multicolumn{4}{c}{{\bf Uranium}}\\
\hline
                  & high $a_{cc}$  & low $a_{cc}$  & depleted upper mantle \\  
\hline
$M$(crust)      & 0.42       & 0.21 	& 0.42 \\
$M$(upper mantle) & 0.06	& 0.09 &  0 \\
$M$(transition+lower mantle) &  0.36 & 0.54  & 0.42 \\
\hline
$\Phi$(crust)        & 2.1 	& 1.0 	&  2.1 \\
$\Phi$(upper mantle) & 0.21 	& 0.32	& 0 \\
$\Phi$(transition + lower  mantle) & 0.79 & 1.2 & 0.92 \\
$\Phi$(tot)          & 3.1 & 2.5 & 3.0 \\
\end{tabular}
\label{Tabhigh}
\end{table}

\begin{table}
\caption[yyy]{{\bf Flux dependence on the detector location.} Fluxes 
in unit of $10^6 \, cm^{-1} \, s^{-2}$ are calculated for 
$M_{crust}(U)=M_{mantle}(U)=0.42 \cdot 10^{17} \, kg$, 
$Th/U=3.8$ and $K/U=10^4$. The last column is the average flux for
the same values of $M_{crust}(U)$ and $M_{mantle}(U)$ (BSE model).}
\begin{tabular}{llll}
  	& Kamioka 	& Gran Sasso 	& average \\
\hline
Uranium & 4.0		& 4.6  		& 3.1 \\
Thorium	& 3.4		& 3.9		& 2.6 \\
Potassium $(\bar{\nu})$& 14		& 17		& 11	\\
Potassium $(\nu)$ 	& 1.8		& 2.1	 	& 1.4\\
\end{tabular}
\label{Tabsites}
\end{table}


\begin{thebibliography}{99}

\bibitem{Bro}
G.C. Brown and A.E. Mussett, ``The Inaccessible Earth'', George Allen \& Unwin, London 1981.


\bibitem{Gal}
GALLEX coll., Phys.Lett. B 447 (1999)158; 

SAGE coll., J. Exp. Theor. Phys. 95 (2002) 181.

GNO coll., Phys. Lett. B 490 (2000) 16.

\bibitem{Ver}
J. Verhoogen, ``Energetics of the Earth'', National Academy of Sciences, 
Washington D.C., 1980.

\bibitem{Bah}
``Solar Neutrinos, the First Thirty Years'', 
edited by  J.N. Bahcall et al., Addison-Wesley New York, 1995.

J.N. Bahcall, ``Neutrino Astrophysics'', Cambridge University Press,
Cambridge 1989.

\bibitem{And}
Landolt-B\"{o}rnstein, ``Numerical data and functional relationships
in science and technology'', New Series, Group IV vol. 3a, Springer-Verlag, Berlin 1993.
http://www.landolt-boernstein.com/; http://ik3frodo.fzk.de/beer/pub/PB\underline{ }198-203.pdf.

E. Anders and N. Grevesse, Geoch. Cosmoch. Acta, 53 (1989) 197.

\bibitem{Sno}
SNO coll., Phys. Rev. Lett. 89 (2002) 011301


\bibitem{Kam}
J. Busenitz, Int. J. Mod. Phys. A 16 S1B, (2001) 742-744.
 
KamLAND Collaboration  May 2002,  hep-ex/0205041.


\bibitem{Bor}
  A.Derbin, O.Smirnov for the Borexino collaboration,     
Phys.Lett. B525 (2002) 29-40
   

\bibitem{Eder}
G. Eder, Nucl. Phys. 78 (1966) 657

\bibitem{Krauss}
L.M. Krauss, S.L. Glashow and D.N. Schramm,
Nature 310 (1984) 191.

\bibitem{Rag}
R.S. Raghavan et al. , Phys. Rev. Lett. 80 (1998) 635.

\bibitem{Rot}
C.G. Rothschild, M.C. Chen and F.P. Calaprice, nucl-ex/9710001,
Geophys. Research. Lett. 25 (1998) 1083.


\bibitem{Wedepohl}
K.H. Wedepohl, Geoch. Cosm. Acta 59 (1995) 1217

\bibitem{MacLennan85}
Taylor and McLennan, ``The Continental Crust'', Beacknell Scientific, Oxford, 1985.

\bibitem{Rin}
A. Ringwood, ``Origin of the Earth and Moon'', Springer-Verlag, New York 1979.

\bibitem{Und}
``Understanding the Earth'', edited by G. Brown et al., Cambridge University Press, Cambridge 1992.

\bibitem{Mc}
Mc. Donough and S. Sun, Chem. Geol. 120 (1995) 223.

\bibitem{Fogli}
G. Fogli et al., Phys.Rev. D 66 (2002) 053010.

\bibitem{Lag}
P.O. Lagage, Nature 316 (1985) 420.

\bibitem{Ponte}
B. Pontecorvo, Oil and Gas Journal 40 (1941) 32.

\bibitem{Smirnov}
A. de Rujula, S.L. Glashow and G. Charpak, Phys. Rep. 99 (1983) 342

A. Ioannisian and A. Smirnov, hep-ph/021012.

T. Ohlsson and W. Winter, Europhys. Lett. 60 (2002) 34.

\bibitem{kamland}
K. Heguchi et al., KamLAND coll., Phys. Rev. Lett. 90 (2003) 021802.

\bibitem{noi2}
G. Fiorentini et al., hep-ph/0201042.

\end{thebibliography}
\end{document}